\documentclass[12pt]{article}
\usepackage{epsfig} 
\usepackage{amssymb}

\def\pr#1{#1^\prime}

\def\beq{\begin{equation}}
\def\eeq{\end{equation}}

\def\beqn{\begin{eqnarray}}
\def\eeqn{\end{eqnarray}}
\relax

\newcommand\Ecm{E_{\rm cm}}

\catcode`\@=11

\@addtoreset{equation}{section}
\def\theequation{\thesection.\arabic{equation}}

\def\@normalsize{\@setsize\normalsize{15pt}\xiipt\@xiipt
\abovedisplayskip 14pt plus3pt minus3pt%
\belowdisplayskip \abovedisplayskip
\abovedisplayshortskip \z@ plus3pt%
\belowdisplayshortskip 7pt plus3.5pt minus0pt}

\def\small{\@setsize\small{13.6pt}\xipt\@xipt
\abovedisplayskip 13pt plus3pt minus3pt%
\belowdisplayskip \abovedisplayskip
\abovedisplayshortskip \z@ plus3pt%
\belowdisplayshortskip 7pt plus3.5pt minus0pt
\def\@listi{\parsep 4.5pt plus 2pt minus 1pt
     \itemsep \parsep
     \topsep 9pt plus 3pt minus 3pt}}
\@twosidetrue

\catcode`@=12

\newlength{\capwidth}
\catcode`\@=11

\def\section{\@startsection{section}{1}{\z@}{3.5ex plus 1ex minus
   .2ex}{2.3ex plus .2ex}{\large\bf}}

\def\thesection{\arabic{section}}

\def\appendix{\setcounter{section}{0}
 \def\thesection{Appendix \Alph{section}:}
 \def\theequation{\Alph{section}.\arabic{equation}}
  }

\catcode`\@=12

\def    \be             {\begin{equation}}
\def    \ee             {\end{equation}}
\def    \ba             {\begin{eqnarray}}
\def    \ea             {\end{eqnarray}}

\def    \=              {\;=\;}
\def    \frac           #1#2{{#1 \over #2}}

\def \as   {\alpha_{\rm s}}

\def \pt   {p_{\rm\scriptscriptstyle T}}

\def \mt   {\ifmmode m_{\rm \scriptscriptstyle T}
                \else $m_{\rm \scriptscriptstyle T}$ \fi}

\def \mur  {\mu_{\rm \scriptscriptstyle{R}}}
\def \muf  {\mu_{\rm \scriptscriptstyle{F}}}

\def \to   {\mbox{$\rightarrow$}}
\newcommand     \MSB            {\ifmmode {\overline{\rm MS}} \else
                                 $\overline{\rm MS}$  \fi}
\newcommand\ssrm{\scriptscriptstyle\rm}
\newcommand\nf{n_{\rm f}}
\newcommand\nlf{n_{\rm lf}}
\newcommand\Tf{T_{\ssrm F}}

\newcommand\Cf{C_{\ssrm F}}

\newcommand\aem{\alpha_{\rm em}}
\newcommand\FO{\rm FO}
\newcommand\RS{\rm RS}
\newcommand\FOMZ{\rm FOM0}
\newcommand\FOpnt{\rm FO_{pnt}}
\newcommand\FOhdr{\rm FO_{hdr}}
\newcommand\RSpnt{\rm RS_{pnt}}
\newcommand\RShdr{\rm RS_{hdr}}
\newcommand\FOMZpnt{\rm FOM0_{pnt}}
\newcommand\FOMZhdr{\rm FOM0_{hdr}}

\begin{document}
\begin{titlepage}
\nopagebreak
{\flushright{
        \begin{minipage}{4cm}
	Bicocca-FT-01-01\hfill\\
	GEF/TH-2-01\hfill\\
	YITP-SB-01-01\hfill\\
        hep-ph/0102134\hfill \\
        \end{minipage}        }

}
\vfill
\begin{center}
{\LARGE 
{ \bf \sc
The $p_{\scriptscriptstyle \rm T}$ Spectrum in Heavy-Flavour
Photoproduction }}
\vskip .5cm

{\large Matteo Cacciari$^1$, Stefano Frixione$^2$ and Paolo Nason$^3$}

\vskip .5cm
{\sl
$^1$C.N. Yang Institute for Theoretical Physics\\
State University of New York\\
Stony Brook, NY 11794-3840\\
\vskip .5cm

$^2$INFN, Sezione di Genova\\
Via Dodecaneso 33, 16146 Genova, Italy\\
\vskip .5cm

$^3$INFN, Sezione di Milano\\
Via Celoria 16, 20133 Milan, Italy
}

\end{center}
\nopagebreak
\vfill
\begin{abstract}
  We consider the transverse-momentum distribution
  of heavy flavours in photon-hadron collisions. 
  We present a formalism in which large transverse-momentum
  logarithms are resummed to the next-to-leading level, and
  mass effects are included exactly up to order $\aem\as^2$, so as
  to retain predictivity at both small and large
  transverse momenta. Phenomenological applications relevant to 
  charm photoproduction at HERA are given.
\end{abstract}
\vskip 1cm
Preprint\hfill \\
January 2001 \hfill
\vfill
\end{titlepage}

\section{Introduction}
This work deals with the computation of the transverse momentum
distribution in heavy flavour photoproduction.
At present, fixed order (FO) calculations are available, including
NLO (Next-to-Leading-Order) corrections~\cite{EllisNasonPh,Smithetal}.
Furthermore, for very large transverse momenta,
the so-called fragmentation function (or resummed) formalism, 
that allows to
resum enhanced terms of order $\aem\as (\as\log\pt/m)^i$
(which we call leading-logarithmic terms, or LL),
plus terms of order $\aem\as^2 (\as\log\pt/m)^i$ (next-to-leading
logarithmic terms, or NLL) has been developed in ref.~\cite{CacciariPhoton},
building upon ref.~\cite{Mele91}.
This approach has however the drawback that it is essentially a ``massless''
formalism, in the sense that it does not include contributions to the
cross section that are suppressed by powers of $m/\pt$.

Several H1~\cite{H1charm} and ZEUS~\cite{ZEUScharm} results are
presented in comparison either with calculations performed in the resummed
(massless) approach (RS), or with the fixed-order NLO calculation.
It is thus important to provide a framework
for computing a heavy-flavour cross section which is accurate
in both the large and the small transverse momentum
regions.  A method for performing the merging of the 
FO~\cite{Nason88,Nason89,Beenakker89,Beenakker91} and RS~\cite{Cacciari94}
calculations in the hadroproduction
case was developed in ref.~\cite{cgn}. In the hadroproduction context
it was found that the mass corrections are positive and large,
a result somewhat contrary to the intuitive belief that masses
reduce the phase space, and thus the cross sections.

The aim of the present work is to extend the formalism of ref.~\cite{cgn}
to the photoproduction case. This extension is not a straightforward one.
In fact, it is well known that a generic
photoproduction cross section has to be written as the sum of
two components (\emph{pointlike} and \emph{hadronic},
also called \emph{direct} and \emph{resolved} respectively).
For the transverse momentum spectrum of a heavy quark we write
\beq
\frac{d\sigma}{dy\,d\pt^2}=\left.\frac{d\sigma}{dy\,d\pt^2}\right|_{\rm pnt}
+ \left.\frac{d\sigma}{dy\,d\pt^2}\right|_{\rm hdr}\,,
\label{phplushd}
\eeq
where pnt stands for pointlike and hdr for hadronic; $y$ is the rapidity
of the heavy quark in the laboratory frame, and
\beqn
\left.\frac{d\sigma}{dy\,d\pt^2}\right|_{\rm pnt}&=&\sum_j\int dx_{\rm p}\,
F^{(H)}_j(x_{\rm p})\,\frac{d\hat{\sigma}_{\gamma j}}{dy\,d\pt^2}
(P_\gamma,x_{\rm p}P_H),
\label{phxsec}
\\
\left.\frac{d\sigma}{dy\,d\pt^2}\right|_{\rm hdr}&=&
\sum_{ij}\int dx_\gamma\,dx_{\rm p}\,
F^{(\gamma)}_i(x_{\gamma})\,F^{(H)}_j(x_{\rm p})\,
\frac{d\hat{\sigma}_{ij}}{dy\,d\pt^2}(x_{\gamma} P_\gamma,x_{\rm p} P_H).
\label{hdxsec}
\eeqn
The sums run over parton flavours, $P_\gamma$ and $P_H$ are the 
four-momenta of the incoming photon and hadron respectively, and
$d\hat{\sigma}_{\gamma j}$, $d\hat{\sigma}_{ij}$ are the subtracted
partonic cross sections. The pointlike and hadronic components
of eqs.~(\ref{phxsec}) and~(\ref{hdxsec}) are strictly related
beyond the leading order in perturbation theory: none of them
is a physical quantity, only their sum 
(eq.~(\ref{phplushd})) is measurable. This is the origin of the
most serious problem we face when extending the formalism of
ref.~\cite{cgn} to the present case. In this work we 
shall follow the strategy adopted in that paper, but we shall
point out the major differences with respect to it,
due to the problems inherent to eq.~(\ref{phplushd}).

In perturbation theory at next-to-leading order (i.e., the highest
accuracy reached so far in the computation of heavy flavour cross
sections), we have the expansions
\beqn
d\hat{\sigma}_{\gamma j}&=&\aem\as d\hat{\sigma}_{\gamma j}^{(0)}+
\aem\as^2 d\hat{\sigma}_{\gamma j}^{(1)}\,,
\label{phexp}
\\
d\hat{\sigma}_{ij}&=&\as^2 d\hat{\sigma}_{ij}^{(0)}+
\as^3 d\hat{\sigma}_{ij}^{(1)}\,.
\label{hdexp}
\eeqn
However, the parton densities in the photon $F^{(\gamma)}_i$ behave
asymptotically (i.e., at large scales) as $\aem/\as$. Thus,
at least at the formal level, the perturbative expansions
of the pointlike and hadronic components of eqs.~(\ref{phxsec}) 
and~(\ref{hdxsec}) are both series in $\aem\as^k$, as in eq.~(\ref{phexp}).
This allows us to simplify substantially our presentation;
in what follows, we shall write the physical cross section
of eq.~(\ref{phplushd}) as an expansion in $\aem\as^k$. The
reader must keep in mind that, in doing this, we are not 
referring to the pointlike component only, but to the observable
that is actually measured in experiments. When we shall deal
either with the pointlike or with the hadronic component only, we shall
indicate it explicitly.

Having clarified this point, we proceed in our program of 
extending the formalism of ref.~\cite{cgn} to photoproduction
reactions. This means that we shall implement a computation
with the following features:
\begin{itemize}
\item
All terms of order $\aem\as$ and $\aem\as^2$ are included exactly,
including mass effects;
\item
All terms of order $\aem\as\left(\as\log\pt/m\right)^i$
and $\aem\as^2\left(\as\log\pt/m\right)^i$
are included, with the possible exception of
terms that are suppressed by powers of $m/\pt$.
\end{itemize}
To be more specific, let us write schematically the result of
the NLO calculation of the photoproduction cross section as
\begin{equation}
  \frac{d\sigma}{dy\,d\pt^2}=A(m)\aem\as+B(m)\aem\as^2
  +{\cal O}(\aem\as^3)\;.
\label{NLOsch}
\end{equation}
The explicit dependence of $A$ and $B$ upon $\Ecm$ (the centre-of-mass energy),
$y$, $\pt$ and the factorization/renormalization scale $\mu$ is not indicated,
and $\as=\as(\mu)$.
The NLL resummed cross section is given by
\begin{eqnarray}
\frac{d\sigma}{dy\,d\pt^2}&=&
        \aem\as\sum_{i=0}^\infty a_i (\as\log \mu/m)^i
   +    \aem\as^2\sum_{i=0}^\infty b_i (\as\log \mu/m)^i 
\nonumber \\&&
+ {\cal O}(\aem\as^3(\as\log \mu/m)^i) + {\cal O}(\aem\as\times\mbox{PST})\;,
\label{NLLsch}
\end{eqnarray}
where PST stands for terms suppressed
by powers of $m/\pt$ in the large-$\pt$ limit
(possibly with further powers of mass logarithms).
The coefficients $a_i$ and $b_i$ depend upon $\Ecm$, $y$, $\pt$ and $\mu$.
If $\mu \approx \pt$,
they do not contain large logarithms of the order of $\log \pt/m$.
The only large logarithms are the ones explicitly indicated.
Our approach combines the results of eqs.~(\ref{NLOsch}) and (\ref{NLLsch}),
giving
\begin{eqnarray}
  \frac{d\sigma}{dy\,d\pt^2}&=& A(m)\aem\as+B(m)\aem\as^2+
\nonumber \\&&
      \left(  \aem\as\sum_{i=2}^\infty a_i (\as\log \mu/m)^i
 + \aem\as^2\sum_{i=1}^\infty b_i (\as\log \mu/m)^i  \right) \times G(m,\pt)
\nonumber \\&&
+ {\cal O}(\aem\as^3(\as\log \mu/m)^i) 
+ {\cal O}(\aem\as^3\times\mbox{PST}) \;,
\label{FONLLsch}
\end{eqnarray}
where the function $G(m,\pt)$ is quite arbitrary, except that it must
be a smooth function (also in the $\pt\to 0$ limit), and that it must
approach one when $m/\pt\,\to\,0$, up to terms suppressed by powers of
$m/\pt$. Observe that the sums now start from $i=2$ and $i=1$,
respectively, in order to avoid double counting.  Thus, this formalism
contains all the information coming from the fixed-order NLO
calculation {\em and} from the NLL resummed calculation.
The arbitrariness in the function $G$ arises from the fact that we do not
know the structure of power-suppressed terms in the higher orders of the 
NLL resummed calculation. The choice of the function $G$ only affects terms
of order $\aem\as^3$, so far unknown.

Before turning to the practical implementation of eq.~(\ref{FONLLsch}),
we stress that it is important that both the RS and the FO 
approaches are expressed in the same renormalization scheme.
The commonly used FO approach uses a renormalization and factorization
scheme in which the heavy flavour is treated as heavy.
Thus, if we are dealing with charm, we
use $\as$ of 3 light flavours as our running coupling constant,
and the appropriate structure functions should not include the charm
quark in the evolution. The RS approach, on the other hand, also includes
the heavy flavour as an active, light degree of freedom.
This problem can be easily overcome by
a simple change of scheme in the FO calculation. 
Section~\ref{sec:scheme} contains the details of this procedure.

Once this is done, the FO calculation matches exactly the terms
up to order $\aem\as^2$ in the resummed approach, in the limit where
power-suppressed mass terms are negligible. In order to subtract from the RS 
result the terms already present in the FO, we must provide
an approximation to the latter where terms suppressed by powers of the mass
are dropped. We shall call FOM0 this ``massless limit''.
In the simplified notation of eqs.~(\ref{NLOsch}) and (\ref{NLLsch})
we have
\begin{equation}
  A(m)=a_0+\mbox{PST}\,,\quad B(m)=a_1\log\mu/m+b_0+\mbox{PST}\;,
\end{equation}
and the FOM0 approximation is given by
\begin{equation}
  \left.\frac{d\sigma}{dy\,d\pt^2}\right\vert_{\rm FOM0}
             =a_0\aem\as+\left(a_1\log\mu/m+b_0\right)\aem\as^2\;.
\label{FOM0sch}
\end{equation}
Our final result will be given by
\begin{equation}
  \label{eq:merge}
  \mbox{FONLL}=\mbox{FO}\;+\left( \mbox{RS}\; -\; \mbox{FOM0}\right)\;
\times G(m,\pt)\;.
\end{equation}
The notation FONLL stands for fixed-order plus next-to-leading logs.
Formula (\ref{eq:merge}) is our practical implementation
of eq.~(\ref{FONLLsch}).

The quantities appearing in the RHS of eq.~(\ref{eq:merge}) are 
available as Fortran computer codes. Actually, due to the non-physical
splitting of photoproduction cross sections as given in eq.~(\ref{phplushd}),
the pointlike and hadronic components are usually computed by different
packages. It is therefore useful to write
\beq
\FO=\FOpnt+\FOhdr,\;\;\;\;
\FOMZ=\FOMZpnt+\FOMZhdr,\;\;\;\;
\RS=\RSpnt+\RShdr.
\eeq
The package that computes $\FOhdr$ was originally developed in 
ref.~\cite{Nason89}, and subsequently modified in ref.~\cite{cgn},
in order to implement $\FOMZhdr$ (whose analytical form was obtained
in ref.~\cite{Nason89}), and in order to use the appropriate renormalization 
and factorization scheme. $\FOpnt$  was
computed in refs.~\cite{EllisNasonPh} and \cite{Smithetal}, and
$\FOMZpnt$ was obtained in the present work.
The code for $\FOpnt$ (taken from ref.~\cite{EllisNasonPh})
has been extensively modified in the context of the present 
work in a manner analogous to what done for $\FOhdr$ in ref.~\cite{cgn}.
The package that computes $\RShdr$ was presented in ref.~\cite{Aversa}, 
suitably modified for heavy quark fragmentation in ref.~\cite{Cacciari94}, 
and used in ref.~\cite{cgn}.  Finally, the package relevant to $\RSpnt$ 
was written by the authors of ref.~\cite{Aurenche87}, and adapted to heavy 
quark fragmentation in ref.~\cite{CacciariPhoton}. It has undergone some 
further modifications during the course of this work.
It should be clear that FO, FOM0, and RS are
all strictly interrelated. This fact provides us with a way not only 
to test that the various codes mentioned above are mutually consistent,
but also that their implementation in our formalism, 
eq.~(\ref{FONLLsch}), has been carried out correctly. In particular, the
relation between FO and FOM0, and between FOM0 and RS, will be 
the argument of sections~\ref{sec:FOM0} and~\ref{sec:matching} 
respectively. In the latter section, we shall show in particular 
that the FOM0 and RS results differ only by terms of order $\aem\as^3$. 

The paper is organized as follows. In sect.~\ref{sec:scheme}
we describe the procedure to adopt in order to translate the FO
result from a scheme with $\nf\,-\,1$ light flavours to a scheme with
$\nf$ light flavours. In sect.~\ref{sec:FOM0} we give a few details
concerning the calculation of the massless limit of the FO calculation.
In sect. \ref{sec:matching} we check the matching between the FOM0 and
the RS calculation. Unlike in the hadroproduction case, subtleties arise
here due to a different separation of the hadronic and pointlike
contributions in the FO and RS approaches.
We shall see that only in the full (i.e. pointlike plus hadronic)
cross section we have complete matching.
In sect.~\ref{sec:powereff} we examine the size of power-suppressed
effects in order to understand at which value of $m/\pt$
the massless approach gives a sensible approximation to the
massive calculation. The function $G(m,\pt)$ will be chosen on the
basis of the considerations given in this section.
In sect.~\ref{sec:pheno} we describe our full result, for the case
of charm production at HERA.
Finally, in sect.~\ref{sec:conclusions} we give our conclusions.

\section{The change of scheme}
\label{sec:scheme}

As shown in detail in ref.~\cite{cgn}, a change of scheme
from the one with $\nlf$ to the one with $\nf=\nlf+1$ light flavours
brings about the following changes in $\as$ and in the parton densities
\begin{eqnarray} \label{aschangescheme}
  \as^{(\nlf)}(\mur)&=&\as^{(\nf)}(\mur)
        -\frac{1}{3\pi}\Tf \log\frac{\mur^2}{m^2} {\as^{(\nf)}}^2(\mur)
            +{\cal O}(\as^3)\,,
 \\ \label{fgchangescheme}
F^{(\nlf)}_g(\muf)&=&F^{(\nf)}_g(\muf)\left[1+\frac{\as^{(\nf)}(\mur)
\Tf}{3\pi}\log\frac{\muf^2}{m^2}\right]+{\cal O}(\as^2).
\end{eqnarray}
All other parton densities are affected at higher orders in $\as$.
Eqs.~(\ref{aschangescheme}) and~(\ref{fgchangescheme}) are universal,
that is, process-independent. It is easy to convince oneself that the
pointlike and hadronic components of the cross sections transform
independently
under this change of scheme. Thus, for what concerns the latter component,
we can safely use the formulae of ref.~\cite{cgn}. In the pointlike
component, the only effect at ${\cal O}(\aem\as^2)$ is generated by the
Born-level $\gamma g$ cross section. It is a matter of trivial algebra
to conclude that, in order to go from the $\nlf$-flavour to the $\nf$-flavour
scheme,  the fixed-order cross section has to be modified by adding a term
\begin{equation}
\delta \sigma_{\gamma g} = 
-\as \frac{1}{3\pi}\Tf\log\frac{\mur^2}{\muf^2}\; \sigma^{(0)}_{\gamma g} \;
\end{equation}
to the $\gamma g$ cross section. For any reasonable range
of scales, this correction is not large, and it vanishes
for $\muf=\mur$.

In the following, we shall always refer to the FO and FOM0 calculations
performed in the $\nf$-flavour scheme. We shall thus always
assume that $\as$ and the parton densities $F_j$ 
refer to $\as^{(\nf)}$ and $F_j^{(\nf)}$.

\section{Massless limit of the fixed-order calculation}
\label{sec:FOM0}

As in the case of the change of scheme, the massless limit of FO can
clearly be performed independently for the pointlike and hadronic
components. The latter has been considered in ref.~\cite{cgn}; here,
we only deal with the pointlike part.

The massless limit of the fixed-order pointlike cross section formulae 
(in the sense of eq.~(\ref{FOM0sch})) is obtained via algebraic 
methods from the results of ref.~\cite{EllisNasonPh}. As pointed out
in ref.~\cite{Nason89},
the limiting procedure is non-trivial. In fact, the partonic cross
sections at order $\aem\as^2$ contain distributions,
such as delta functions or principal value singularities.
When taking the massless
limit, new contributions to these distributions arise.
We have computed this limit analytically by a computer-algebra procedure
applied to the massive cross section formula, and checked its
correctness in the following
way. We compute the heavy-flavour differential cross section at fixed
$\pt$, $y$, and centre-of-mass energy. We choose the renormalization
and factorization scales equal to $\pt$. Under these conditions,
the mass dependence of the result is confined to the partonic cross
sections. In the massless limit approximation, the only remnants
of mass dependence are in logarithms of the mass in the ${\cal O}(\aem\as^2)$
terms. Thus, if we plot the $\FOMZpnt$ cross section versus the logarithm of
the mass, we get a straight line. On the other hand, if we plot the full
$\FOpnt$ cross section versus the logarithm of the mass,
it should approach the $\FOMZpnt$ result in the limit of small masses.
We have performed this test, choosing the proton energy $E_{\rm p}=820\;$GeV
and the photon energy $E_\gamma=20\;$GeV.
We adopt the CTEQ4M set \cite{cteq4m} for the
proton parton density functions. These parameters will be our reference
choice from now on. The results are displayed in figs.~\ref{fig:lim_all}
and \ref{fig:lim_pgpq}.
\begin{figure}[t]
  \begin{center}
    \epsfig{figure=lim_all.eps,width=10cm}
\parbox{\capwidth}{
    \caption{\label{fig:lim_all} \protect\small
       Comparison of the $\FOpnt$ and $\FOMZpnt$ differential 
       cross sections as a function of the logarithm of the mass,
       at $\pt=20$ GeV and $y=1$.}
}
  \end{center}
\end{figure}
\begin{figure}[t]
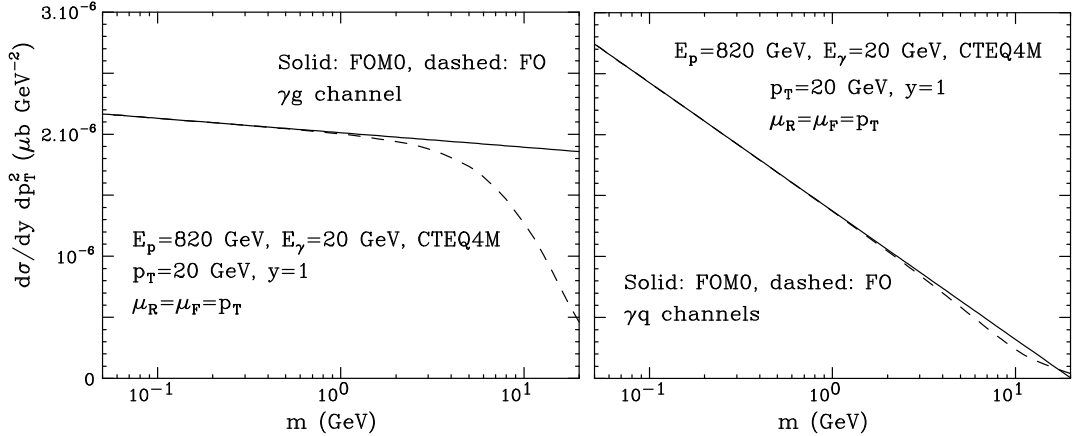

  \begin{center}
    \epsfig{figure=lim_pg.eps,height=5.8cm}
    \epsfig{figure=lim_pq.eps,height=5.66cm}
\parbox{\capwidth}{
    \caption{\label{fig:lim_pgpq} \protect\small
       As in figure~\ref{fig:lim_all}, for $\gamma g$ (left)  and $\gamma q$
        (right) components alone.}
}
  \end{center}
\end{figure}
From the figures, it is quite apparent that the massless limit,
as well as its implementation for the calculation of
cross sections, was carried out correctly.
There is also an important observation to make: the $\FOMZpnt$
cross section is {\em larger} than the massive calculation, i.e. power
suppressed mass effects are negative, 
contrary to the case of hadroproduction.

Notice also that the $\FOMZpnt$ approximation is quite accurate even at
relatively large values of $m/\pt$. For example, from both
figs.~\ref{fig:lim_all} and~\ref{fig:lim_pgpq} we notice that even for
$m/\pt\approx 1/2$, the $\FOpnt$ cross section differs from $\FOMZpnt$
by at most 30\%. This has to be contrasted with the case of the
hadronic component (see ref.~\cite{cgn}), where $\FOMZhdr$ is 
describing well $\FOhdr$ only for rather small values of $m/\pt$.
This mismatch between the behaviour of the pointlike and the hadronic
component will clearly show up in phenomenologically relevant cases,
as we shall later see in sect.~\ref{sec:pheno}.

\section{Matching}
\label{sec:matching}

We now examine the matching between the resummed approach and the FOM0
calculation. In this case, there is a strict interplay between
the pointlike and the hadronic components, that should be dealt with
very carefully.

There are ingredients in the resummed approach that are not explicitly
present in the FOM0 calculation. These are the fragmentation functions
for final-state partons to go into the heavy quark, and the parton density
for finding a heavy quark inside the hadron.
The fragmentation function for any parton
to go into a heavy quark has a power expansion in terms
of the coupling constant evaluated at the scale $\mu$\footnote{We take
for simplicity $\mur=\muf=\pt$, and denote the common value with $\mu$.},
and of logarithms of $\mu/m$:
\begin{equation}
  \label{eq:Dseries}
  D_j(x,\mu,m)=\sum_{k=0}^\infty \sum_{l=0}^k
         d_j^{(k,l)}(x)\,\log^l\frac{\mu}{m}\,\as^k(\mu)\,,
\end{equation}
that can be obtained by solving the evolution equation for the fragmentation
function at the NLL level, with the initial conditions
of ref.~\cite{Mele91}.
Similarly, the parton density for finding
the heavy flavour in a hadron can be expanded in the form
\begin{equation}
  \label{eq:Fseries}
  F_h(x,\mu,m)=\sum_{k=0}^\infty
 \sum_{l=0}^k f^{(k,l)}(x,F_l(\mu))\log^l\frac{\mu}{m}\,\as^k(\mu)\;.
\end{equation}
With $F_l(\mu)$ in the argument of the coefficients, we mean that the 
coefficients have a complicated {\em functional} dependence upon the parton
densities evaluated at the scale $\mu$.
The existence of formal expansions of the form~(\ref{eq:Dseries})
and (\ref{eq:Fseries}) can be easily proved, by writing the Altarelli--Parisi
equations in integral form, and then solving them iteratively.
A more detailed argument was given in Appendix A of ref.~\cite{cgn}.

Once eqs.~(\ref{eq:Dseries}) and (\ref{eq:Fseries}) are formally
substituted in the RS cross section formula,
this formula itself becomes a power expansion of the form
of eq.~(\ref{NLLsch}), with the coefficients that depend
(functionally) upon the structure functions for light partons,
in the $\nf$-flavours scheme,
evaluated at the scale $\mu$. The FOM0 calculation has an expansion
of the same form (truncated to order $\aem\as^2$) with coefficients
that are also dependent upon the same light-parton structure
functions\footnote{We observe that this property of the FOM0
calculation is only valid in the modified scheme described in section
\ref{sec:scheme}. If we had used the standard scheme for the fixed-order
calculation, the structure functions and the coupling $\as$
appearing there would be those with $\nf-1$ flavours.}
evaluated at the scale $\mu$.
Thus, because of the next-to-leading logarithmic accuracy of the
resummed cross section, the terms up to the order $\aem\as^2$ in $\RS$
will match exactly with the FOM0 calculation.

\begin{figure}[t]
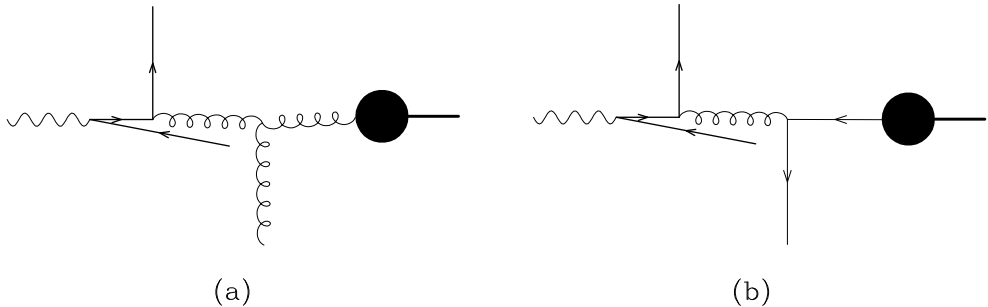

  \begin{center}
    \epsfig{figure=photon-split1.eps,width=6cm}\hskip 1cm
    \epsfig{figure=photon-split2.eps,width=6cm}
\parbox{\capwidth}{
    \caption{\label{fig:photon-split} \protect\small
       Photon splitting into a heavy-quark pair.}}
  \end{center}
\end{figure}

In the photoproduction case, the matching can take place
only in the full cross section, i.e. pointlike plus hadronic.
In fact, let us consider the photoproduction
subprocess in which a photon splits into a heavy-quark pair,
and afterwards the heavy quark scatters with a parton
coming from the hadron, as shown in fig.~\ref{fig:photon-split}.
The contributions depicted in the figure, in the kinematic region
of a photon splitting collinearly, is fully included in
the pointlike contribution in the massive, fixed order  calculation.
In fact, because of the mass, no collinear subtractions are
needed on the photon side.

In the resummed approach, instead, these
graphs are collinear divergent, and the singularities are subtracted
at a scale $\mu$. A corresponding hadronic contribution is present,
where a heavy quark is found inside the incoming photon.
Such a contribution is not present in the hadronic part
of the fixed-order approach. 

Therefore, in order to check the matching, we should
consider pointlike plus hadronic cross sections, and
compute the difference between two relatively large
numbers (RS and FOM0), each of which is obtained as the sum of
two numbers ($\RSpnt$ plus $\RShdr$ and $\FOMZpnt$ plus $\FOMZhdr$).
In practice, this procedure requires an extremely careful treatment
of numerics, which is not required anywhere else in our study. We
thus adopted a slightly different, although equivalent, strategy,
which is based upon the observation that, as far as the matching is
concerned, it is only the contribution depicted in fig.~\ref{fig:photon-split}
that is treated differently in RS and FOM0. Therefore, we can simply
add to the resummed pointlike result
the hadronic contribution with a heavy quark in a photon.
Since the matching is checked up to the order $\aem\as^2$,
it is enough to compute such contribution at this order.
Calling $s$ the $s$-channel invariant of the photon-hadron system
and $y_h$ ($y_l$) the heavy quark (light recoiling parton) rapidity in the
photon-hadron CM system, we immediately find
\begin{equation} \label{photonsplit}
\frac{d\hat\sigma_{\rm \scriptscriptstyle PQ}}{dy_h d\pt^2}=
\int\frac{1}{8\pi s}
\sum_l P_{h\gamma}(x_\gamma)\log\frac{\muf^2}{m^2}
      \;F^{(p)}_l(x_{\rm p},\muf)
\frac{d\hat\sigma^{(0)}_{hl}(\pt,\hat{y})}{d\Phi_2}\; dy_l 
\end{equation}
where PQ stands for photon to quark, and
\begin{equation}
P_{h\gamma}(x)=\frac{\aem}{2\pi}N_c e_h^2 (x^2 + (1-x)^2)\;. 
\end{equation}
$d\hat\sigma^{(0)}_{hl}/d\Phi_2$ is the leading order (i.e. ${\cal O}(\as^2)$)
massless cross section (without the two-body phase space)
for the partons $h$ (the heavy quark) and $l$
(a light quark or a gluon) to give an outgoing heavy quark $h$ with
transverse momentum $\pt$ and rapidity $\hat{y}$ in the parton centre-of mass
(CM). The variables $\pt$ and $\hat{y}$ characterize
completely the scattering kinematics in the partonic CM system. We have
furthermore
\begin{eqnarray}
\hat{y}&=&\frac{y_h-y_l}{2}\,,
\\
x_\gamma &=& \sqrt{\frac{4 \pt^2}{s}}\,\frac{\exp(y_h)+\exp(y_l)}{2}\,, \\
x_{\rm p} &=& \sqrt{\frac{4 \pt^2}{s}}\,\frac{\exp(-y_h)+\exp(-y_l)}{2}
\end{eqnarray}
and the partonic cross sections are
\begin{eqnarray}
\frac{d\hat\sigma^{(0)}_{hq}(\pt,\hat{y})}{d\Phi_2}&=&(4\pi\as)^2\,
\frac{1}{2\hat{s}}\frac{4}{9} \frac{\hat{s}^2+\hat{u}^2}{\hat{t}^2}\,,
 \\
\frac{d\hat\sigma^{(0)}_{hg}(\pt,\hat{y})}{d\Phi_2}&=&(4\pi\as)^2\,
\frac{1}{2\hat{s}}
\left(-\frac{4}{9}\frac{\hat{s}^2+\hat{u}^2}{\hat{s}\hat{u}}
+\frac{\hat{u}^2+\hat{s}^2}{\hat{t}^2}\right)\,,
\end{eqnarray}
where the partonic Mandelstam variables $\hat{s}$, $\hat{t}$ and $\hat{u}$
are easily obtained from $\pt$ and $y$. For future reference,
we shall call PQ (for Photon to Quark) the contribution of
eq.~(\ref{photonsplit}). Such contribution is part of the direct
component of the FOM0 result.

In order to check the matching of the resummed and fixed order
calculation we proceeded in the following way:
\begin{itemize}
\item[(a)]
we computed the $\FOMZpnt$ result in the $\nf$ flavours scheme;
\item[(b)]
we computed the RS result without the intrinsic heavy quark
component in the hadron parton densities, and with a heavy
quark fragmentation function set equal to $\delta(1-z)$ for the
$D_q$ component, and all other components set to zero;
\item[(c)] we computed the LO contribution to the RS result,
with the heavy quark component in the hadron parton densities
set equal to
\begin{equation}
\label{eq:pdfQapprox}
F^{(H)}_h(x,\mu)
=\frac{\as(\mu) \log\mu^2/m^2}{2\pi}
\int^1_x F^{(H)}_g(x/z,\mu) P_{hg}(z)\frac{dz}{z},
\end{equation}
and all other components set to zero;
\item[(d)]
we computed the LO contribution to the RS result, with the
heavy quark density set to zero, and the heavy quark
fragmentation function set to its ${\cal O}(\as)$ value
\begin{eqnarray}
D_h(z,\mu)& \to& {{\alpha_s(\mu) \Cf}\over{2\pi}}\left[
{{1+z^2}\over{1-z}}\left(\log{{\mu^2}\over{m^2}} -2\log(1-z)
-1\right)\right]_+ \label{DQQ} \\ 
D_g(z,\mu) &\to& {{\alpha_s(\mu) \Tf}\over{2\pi}}(z^2 + (1-z)^2)
\log{{\mu^2}\over{m^2}} \label{DgQ} \\
D_i(z,\mu)& \to& 0 \label{DqQ}\,,\quad\mbox{for $i\ne g,h$}\,.
\end{eqnarray}
\end{itemize}
We then verified that the sum of items (b), (c) and (d), plus the ${\rm PQ}$
contribution (eq.~(\ref{photonsplit})) is exactly equal to item (a).
Observe that with this procedure we were able to isolate the terms
of order $\aem\as$ and $\aem\as^2$ in the RS result, and thus,
in order to check the matching, we did
not need to go to the weak coupling limit, as was done in ref.~\cite{cgn}.
The resummed hadronic component, suitably augmented
of the PQ contribution, must then match the fixed order hadronic component
in the massless limit. This follows easily from the work of ref.~\cite{cgn}.

The procedure we carried out, besides convincing us of the correctness of our
theoretical approach, has also served as a test of consistency between the
computer programs of ref.~\cite{Aurenche87} and the fixed order calculation of
ref.~\cite{EllisNasonPh}.

\begin{figure}[tb!]
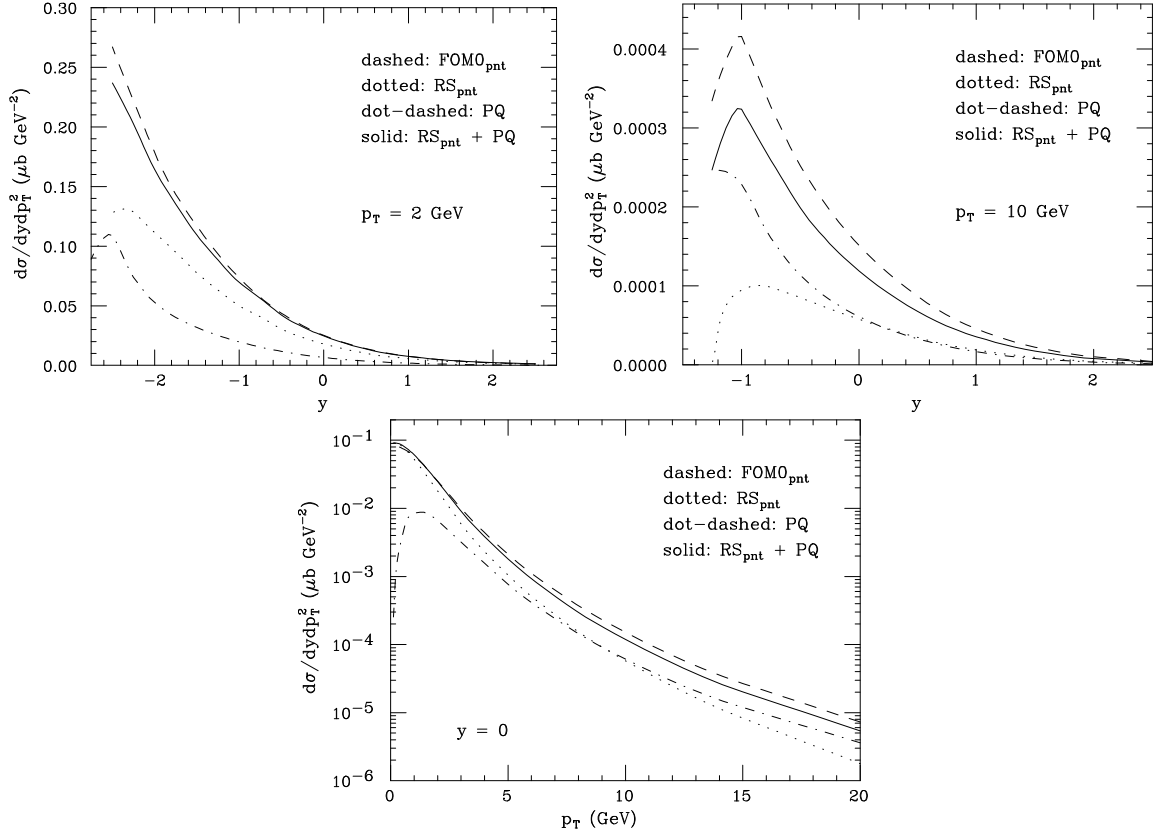

  \begin{center}
    \leavevmode
    \begin{minipage}{7.5cm}
    \epsfig{figure=matching-dir-y-pt2.eps,height=5.5cm}
    \end{minipage}
    \begin{minipage}{7.5cm}
    \epsfig{figure=matching-dir-y-pt10.eps,height=5.5cm}
    \end{minipage}
    \begin{minipage}{7.5cm}
    \epsfig{figure=matching-dir-pt.eps,height=5.5cm}
    \end{minipage}
\parbox{\capwidth}{
\caption{\label{fig:pntmatching} \protect\small $\FOMZpnt$ compared to the sum 
of $\RSpnt$ and PQ. The parameters are as in
fig.~\ref{fig:lim_all}. The charm quark mass was set equal to 1.5~GeV.
}
}
  \end{center}
\end{figure}

We have thus demonstrated that the quantity 
$\RSpnt+{\rm PQ}-\FOMZpnt$ is of order $\aem\as^3$. However, this
does {\em not} guarantee that it is also small in practice.
In ref.~\cite{cgn} it was shown\footnote{See in particular fig. 8
of ref.~\cite{cgn}} that $\RS-\FOMZ$ in the case
of hadronic collisions is of order $\as^4$, but it is numerically 
non-negligible even for bottom production.
We shall now show that, in the present case,
the difference $\RSpnt+{\rm PQ}-\FOMZpnt$ is numerically small,
at least in the region where $\as\log\pt/m$ is also small (that is to say
when $\pt$ is not large).  This is in fact what we
see in fig.~\ref{fig:pntmatching}.
We also see that, as previously discussed, $\FOMZpnt$ is not a good
approximation of $\RSpnt$, due to the lack
of the photon-splitting term ${\rm PQ}$.
We further note that the matching 
deteriorates at higher values of $\pt$, due to the increasing importance 
of the resummation performed in $\RSpnt$ but absent in $\FOMZpnt$.

\section{Power effects in the RS and FOM0 calculations}
\label{sec:powereff}
When comparing and matching the FO, FOM0 and RS approaches, there
is much arbitrariness in the way mass effects are treated.
For example, we may decide to compare transverse-mass distributions
instead of transverse momenta. These are equal for the FOM0 and RS
calculations, but differ in the FO case. In ref.~\cite{cgn}
the effect of this replacement was studied for hadron-hadron cross sections.
Here we thus study only the pointlike component $\FOpnt$
and $\FOMZpnt$.

\begin{figure}[tb!]
  \begin{center}
    \leavevmode
    \epsfig{figure=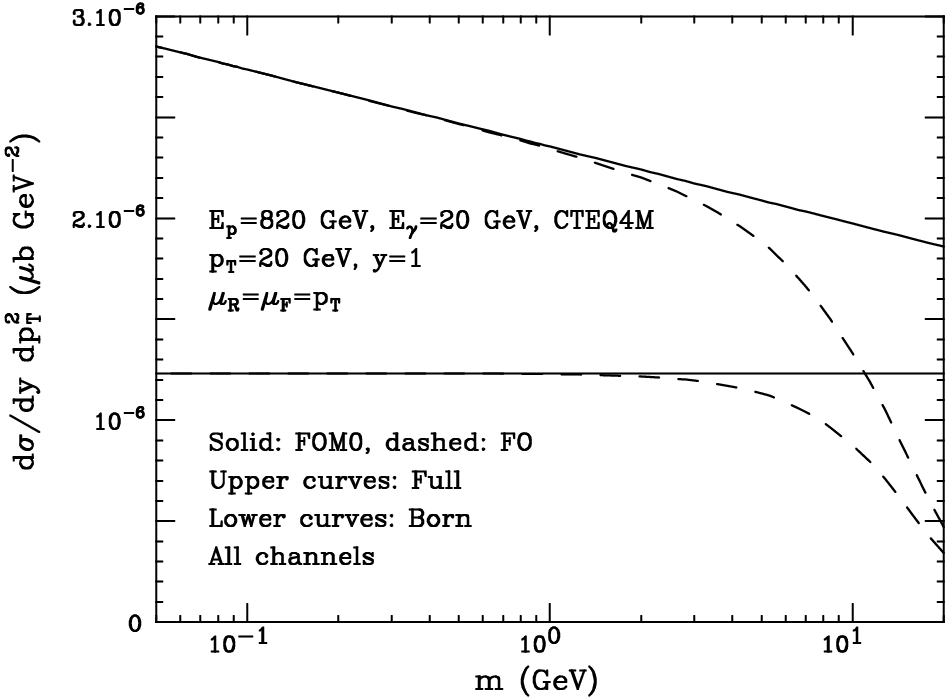,height=5.8cm}
    \epsfig{figure=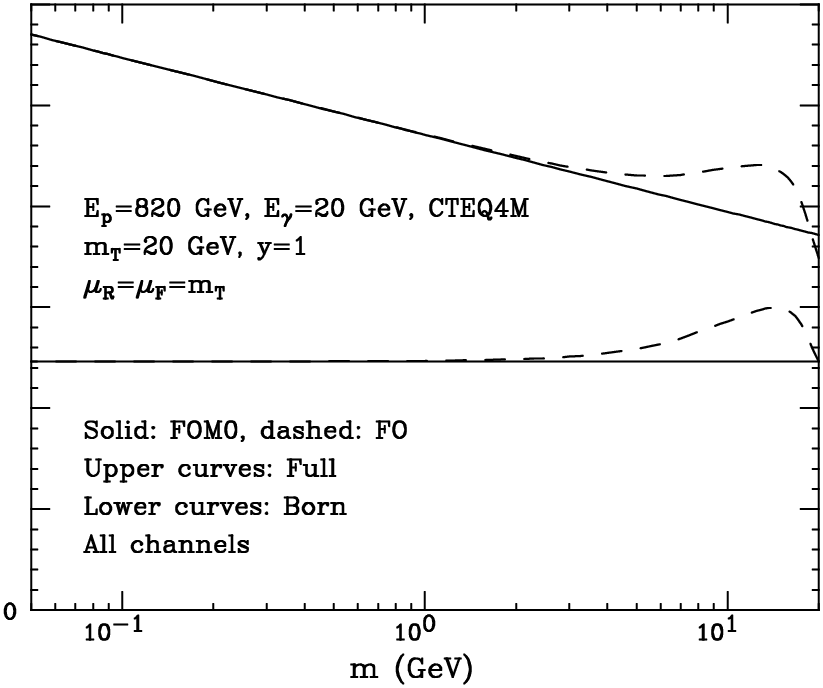,height=5.68cm}
\parbox{\capwidth}{
\caption{\label{fig:pe_plots} \protect\small $\FOpnt$ and
             $\FOMZpnt$ at Born and full ${\cal O}(\aem\as^2)$ level,
             plotted as a function of the mass and at fixed transverse
             momentum (left figure) or fixed transverse mass (right figure).}
}
  \end{center}
\end{figure}

In fig.~\ref{fig:pe_plots} we plot $\FOpnt$ and $\FOMZpnt$
as a function of the mass, keeping either $\mt$ or $\pt$ fixed.
It is clear that, in the plots at fixed transverse momentum,
the massive calculation is more suppressed near the threshold
(i.e., as $m$ approaches $\pt$). In this region higher
$x_{\rm p}$ values are probed in the proton structure function
with respect to the massless calculation since, besides the
transverse momentum, also the mass has to be produced.
This is clearly a spurious effect, and we shall always prefer to
perform the matching at fixed transverse masses. The right plot
of fig.~\ref{fig:pe_plots} is an {\it a posteriori} justification
of this procedure, since it is quite clear that in this case the massive
and massless limit results are much closer to each other. It is also worth
noting that in the right plot of fig.~\ref{fig:pe_plots} power suppressed
mass effects are positive, in contrast to what we observed earlier
(see fig.~\ref{fig:lim_all}).
\begin{figure}[tb!]
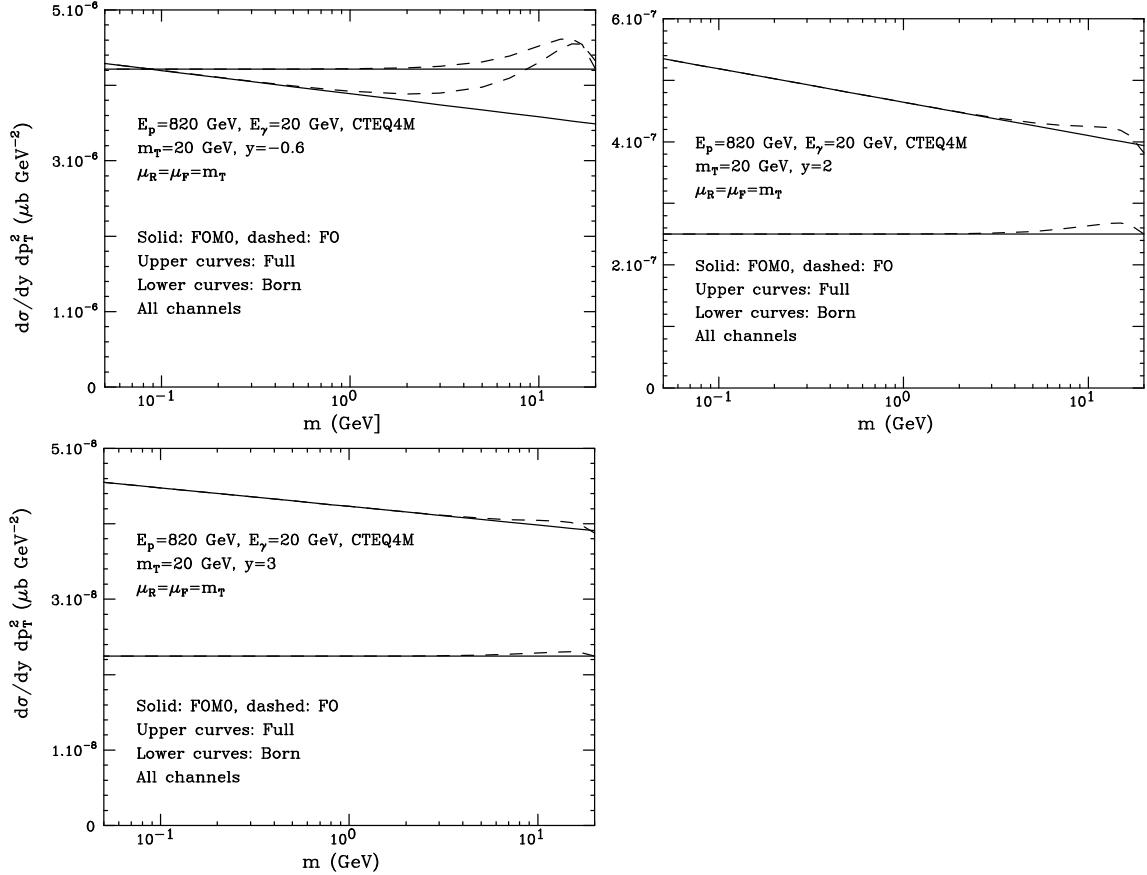

  \begin{center}
    {\leavevmode
    \epsfig{figure=mt_y-06.eps,height=5.8cm}
    \epsfig{figure=mt_y2.eps,height=5.68cm}}
    {\leavevmode
    \epsfig{figure=mt_y3.eps,height=5.8cm}\hskip 7.2cm\phantom{.}}
\parbox{\capwidth}{
\caption{\label{fig:pe_plotsy} \protect\small
             As in the right plot of fig.~\ref{fig:pe_plots},
             for $y=-0.6$, 2 and 3.}
}
  \end{center}
\end{figure}
In figure~\ref{fig:pe_plotsy} we show analogous plots at fixed transverse
mass for $y=-0.6$, 2 and 3. In all cases
the massless limit and full massive results are in good agreement, although
a worsening is seen for the case of $y=-0.6$. This rapidity value
is quite close to the phase space boundary, where we do in fact expect
some anomalous behaviour due to Sudakov logarithms.

We shall now proceed as follows.
For a given transverse momentum $\pt$, the FO cross section is evaluated
and combined, using eq.~(\ref{eq:merge}), with the FOM0 and RS results
evaluated at the corresponding $\mt = \sqrt{\pt^2 + m^2}$ value.
In this way the three calculations are performed at the same $\mt$.
Moreover, a central choice for the renormalization and factorization scales
will be $\mur=\muf=\sqrt{\pt^2 + m^2}$, so that they coincide
in the three calculations.

In the hadroproduction case~\cite{cgn}, a suppression factor $G(m,\pt)$
was introduced to multiply the RS$\,-\,$FOM0 term in eq.~(\ref{eq:merge}),
with $G(m,\pt)$ approaching one at large $\pt$. The following form was chosen:
\begin{equation}
G(m,\pt)=\frac{\pt^2}{\pt^2 + c^2 m^2}\;.
\end{equation}
Adopting the same form, our final formula becomes
\begin{equation}
  \label{eq:merge2}
  \mbox{FONLL}=\mbox{FO}\;+\; {{\pt^2}\over{\pt^2 + c^2 m^2}}
  [\mbox{RS}\;-\; \mbox{FOM0}]\;.
\end{equation}
In ref.~\cite{cgn} $c=5$ was chosen,
in order to suppress meaningless large corrections coming from
the massless approach at low momenta.
A detailed discussion of the role of $c$ in the present case will be given
in the next section.

\section{Phenomenological results}
\label{sec:pheno}

We now present some benchmark results of our calculation, based
upon eq.~(\ref{eq:merge2}). In order to perform the calculation,
we have suitably modified the codes relevant to $\FOpnt$, $\FOMZpnt$,
and $\RSpnt$, in order to obtain a consistent implementation.
The hadronic quantities, on the other hand, have been produced with the
program of ref.~\cite{cgn}, here modified
in order to allow the use of the photon parton densities. 

\begin{figure}[tb!]
  \begin{center}
    \epsfig{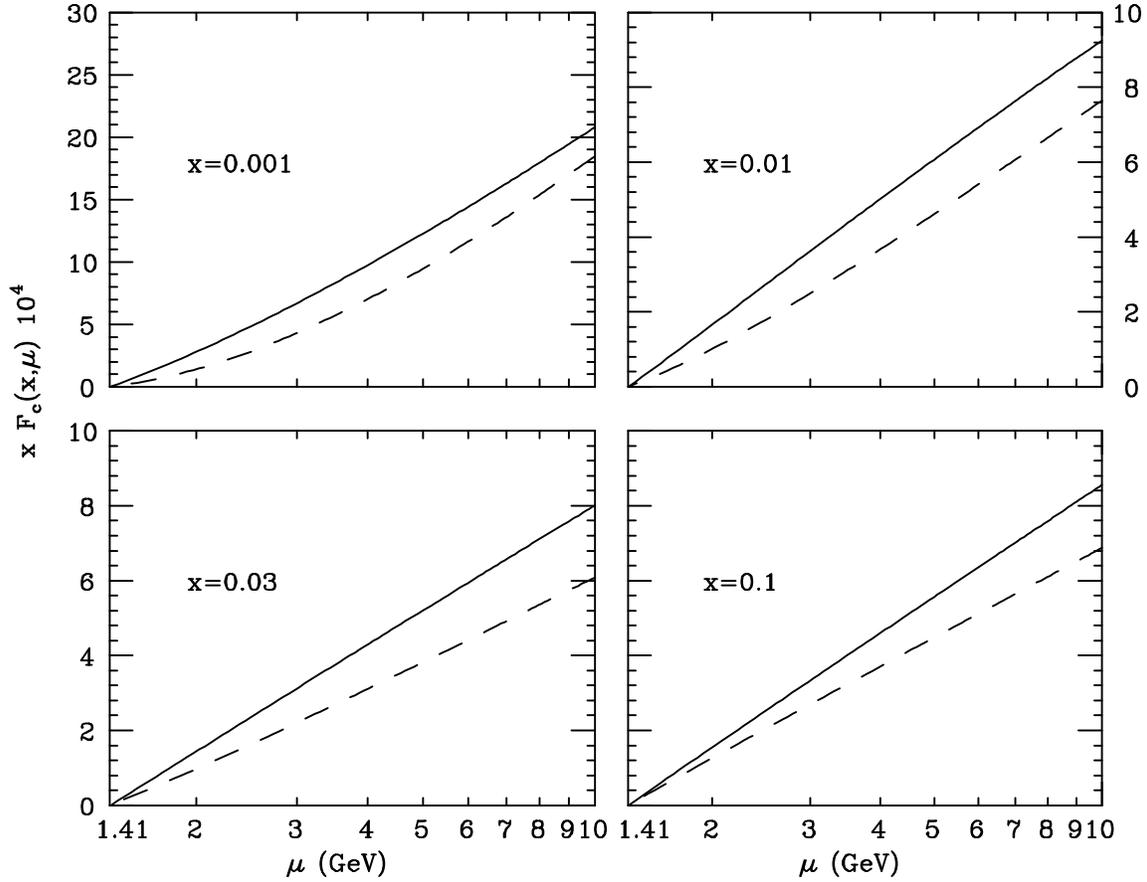}
\parbox{\capwidth}{
\caption{\label{fig:charmpdf} \protect\small
            Charm parton density according to the AFG parametrization
            (solid), compared to eq.~(\ref{cpdfapprox}) (dashed).}
}
  \end{center}
\end{figure}

When applying our matched formalism to phenomenology, we must first
make sure that the parton densities we use correctly incorporate
the logarithms of $m/\mu$ that we are trying to resum. This must
be the case if the structure functions have been evolved correctly,
with the heavy flavour evolution turned on when $\mu=m$, as appropriate
in the \MSB scheme~\cite{Collins86}. In the structure function fits
available today this is not always the case. In part this is due to the fact
that the parton densities are often given as an interpolating grid,
and the region near the heavy flavour threshold may not be represented
accurately. Furthermore some parton density sets do not
implement the heavy flavour thresholds according to ref.~\cite{Collins86}.
While for hadron structure
functions several sets with a correct charm density are available,
the choice among photon structure functions is quite limited.
We have chosen the AFG set~\cite{AFG}, which is in the \MSB scheme, and
claims a correct implementation of the charm density.
In order to test this fact, we have plotted in fig.~\ref{fig:charmpdf}
the AFG charm parton density in the photon together with a charm density
computed with the following formula:
\begin{equation}
\label{cpdfapprox}
F_c(x,\muf)=\left[
\frac{\aem N_c}{2\pi} \, c_c^2(x^2+(1-x)^2)
+ \frac{\as(\muf)\Tf}{2\pi}\int_x^1
(z^2+(1-z)^2) F_g(x/z) \frac{dz}{z} \right] \log\frac{\muf^2}{m^2}\;,
\end{equation}
(where $c_c=2/3$ is the electric charge of the charm quark)
that should hold for factorization scales not too far from the heavy 
quark mass.
In order to match the AFG parameters, we have chosen $m=1.41\;$GeV in this
plot. We expect that the slope of the AFG charm density
and our approximate formula
should agree for small scales. Unfortunately we do not observe this,
especially at small values of $x$. We have checked, however, that these
differences would only be marginally important if our approximate formula,
eq.~(\ref{cpdfapprox}), were
used to calculate physical cross sections, the results obtained with the
original AFG charm densities being well reproduced.

\begin{figure}[tb!]
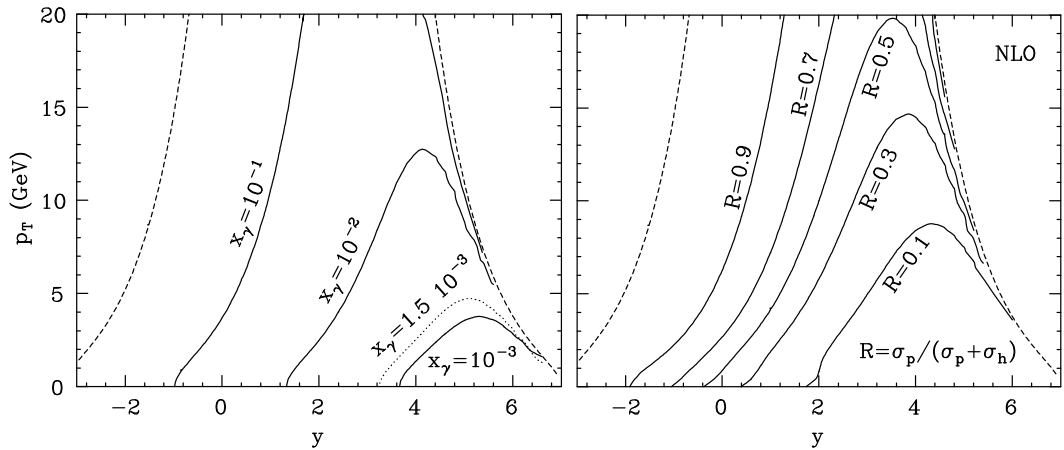

  \begin{center}
    \epsfig{figure=x1min.eps,height=5.9cm}
    \epsfig{figure=pnt_o_tot.eps,height=5.82cm}
\parbox{\capwidth}{
\caption{\label{fig:x1min} \protect\small
            Contours in the $y$-$\pt$ plane of constant minimum $x_\gamma$
            value probed in the hadronic component (on the left),
            and of constant ratio of the direct component over the total
            cross section (on the right) for the NLO
            result. The short--dashed line represents the phase space
            boundary.}
}
  \end{center}
\end{figure}

In what follows, we consider our reference case of photon-proton collisions, 
with $E_\gamma=20\;$GeV and $E_p=820\;$GeV.
In the left panel of fig.~\ref{fig:x1min} we plot contour lines representing
the minimum value of $x_\gamma$ allowed in the hadronic component of the cross
section, in the $y$-$\pt$ plane.
Commonly available photon structure function fits are undefined below values 
of $x_\gamma$ of the order of
$10^{-3}$.
The minimum value of $x_\gamma=1.5\;10^{-3}$ relevant to the AFG set is
explicitly shown in the figure.
For rapidities currently probed at HERA, smaller $x_\gamma$ values are
never reached. In the right panel of fig.~\ref{fig:x1min}
we show, for future reference, the ratio $R$ of the pointlike component
over the total cross section, both computed at the NLO.
Where the hadronic component prevails (i.e. $R$ becomes small),
we expect to find the same problems found in ref.~\cite{cgn} as far as the
matching between RS and FOM0 is concerned. We point out that the results
displayed in the right panel of fig.~\ref{fig:x1min} have a scale dependence 
of ${\cal O}(\aem\as^2)$, and they are thus accurate to LO only; however, the 
plot gives a clear idea on the dominance of pointlike or hadronic components
in the physical cross sections.

\begin{figure}[tb!]
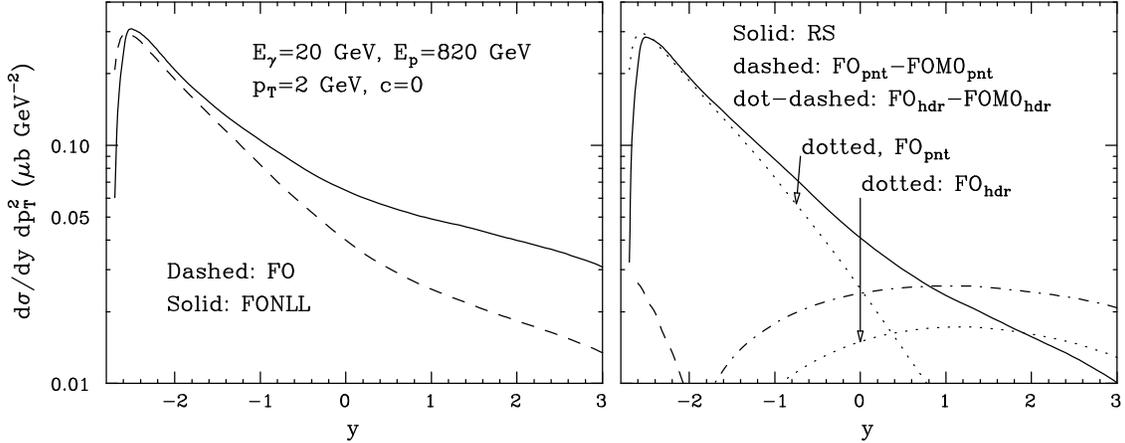

  \begin{center}
    \epsfig{figure=rap-pt2-nosuda.eps,height=5.9cm}
    \epsfig{figure=rap-pt2-nosuda-breakup.eps,height=5.9cm}
\parbox{\capwidth}{
\caption{\label{fig:rap-pt2-nosuda} \protect\small
  Rapidity distribution of charm quarks at $\pt=2\;$GeV, with no smearing
  factor for the term $\mbox{\rm RS}-{\rm FOM0}$ (i.e., $c=0$).
  In the right figures the various contributions to the FONLL cross section
            are shown.
}}
  \end{center}
\end{figure}

In our phenomenological study we shall use $m=1.5\;$GeV,
the AFG set for the photon and CTEQ4M for the proton parton densities.
In fig.~\ref{fig:rap-pt2-nosuda} we show the rapidity distribution
for charm quarks at $\pt=2\;$GeV.
In the left plot of this figure, both the fixed order prediction and
our FONLL results are shown. The photon has negative rapidity in our
convention. The phase space limit for the rapidity at $\pt=2\;$GeV is
$-2.77<y<6.49$. Thus, the left limit of the plots is near the
negative rapidity limit. No suppression factor was applied to the
difference ${\rm RS-FOM0}$. At these low values of $\pt$, we would expect
that the fixed order and the FONLL result should agree. In fact we
observe good agreement at negative rapidity, while at moderate and
large rapidity the FONLL result is much larger than the fixed order
one.

A detailed analysis of the various contributions to the FONLL
cross section is shown in the right plot of the figure.  We show the
RS result, and the mass corrections FO-FOM0,
for the pointlike and hadronic component separately. For comparison, also
the fixed order pointlike and hadronic component are shown.  

\begin{figure}[tb!]
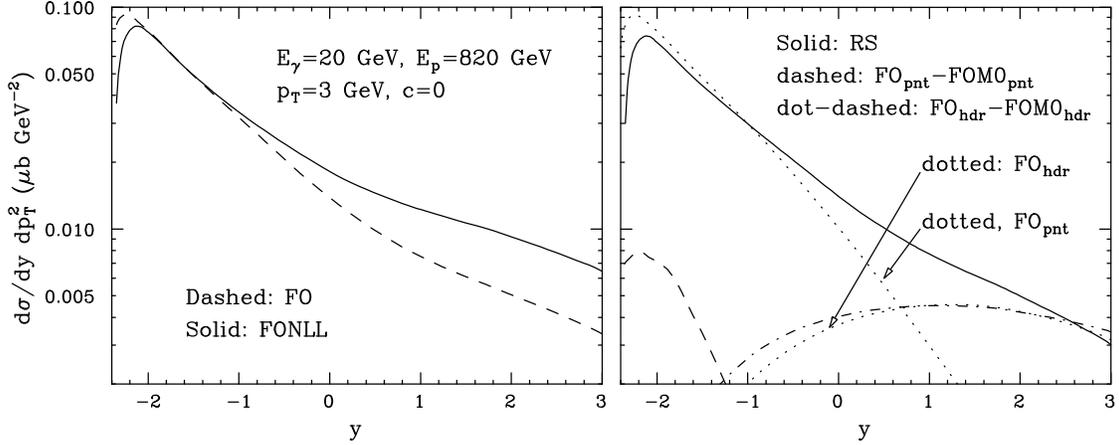

  \begin{center}
    \epsfig{figure=rap-pt3-nosuda.eps,height=5.9cm}
    \epsfig{figure=rap-pt3-nosuda-breakup.eps,height=5.83cm}
\parbox{\capwidth}{
\caption{\label{fig:rap-pt3-nosuda} \protect\small
  As in fig.~\ref{fig:rap-pt2-nosuda}, for $\pt=3\;$GeV.
}}
  \end{center}
\end{figure}

A few comments are in order. We begin by
noticing that the difference $\FOpnt-\FOMZpnt$ (the
dashed line in the left lower corner of the plot) is quite small compared to
$\FOpnt$.  On 
the contrary, $\FOhdr-\FOMZhdr$ is
even larger than the $\FOhdr$ result, due to a negative $\FOMZhdr$
contribution. Comparing the pointlike and hadronic fixed order results, we
see that the first one prevails up to rapidities of 0.4.
A similar plot, for $\pt=3\;$GeV, is shown in fig.~\ref{fig:rap-pt3-nosuda}.
We see similar features to the case of fig.~\ref{fig:rap-pt2-nosuda}.

We conclude
that the large difference between our FONLL and FO results
is only due to the hadronic component of the cross
section. 

This is consistent with what already observed in
ref.~\cite{cgn}.
In that paper these differences were ascribed to a large ${\rm RS-FOM0}$ term
which, due to spurious higher order terms present in RS, did not 
vanish quickly enough in the small $\pt$ limit. Consequently, 
the small-$\pt$ suppression factor $G(m,\pt)=\pt^2/(\pt^2+c^2 m^2)$, with 
$c=5$, was applied to the difference ${\rm RS-FOM0}$ in
order to get rid of this problem. 

In this work we can now make a
conjecture about a possible relation between the higher order behaviour of the
``massless'' RS calculation and the size of the mass terms: in the hadronic case
mass terms are large and so are the higher orders in RS. In the pointlike 
component, instead, the mass terms are relatively small, and there appears
to be a good cancellation between RS and FOM0. One can therefore make the
reasonable assumption that the less important the power suppressed mass terms
are, the better behaved a resummed ``massless'' approximation will be.

Having said so, we observe that for the pointlike component alone no
suppression factor is actually needed, consistently with what we
observed in section~\ref{sec:matching}.

\begin{figure}[tb!]
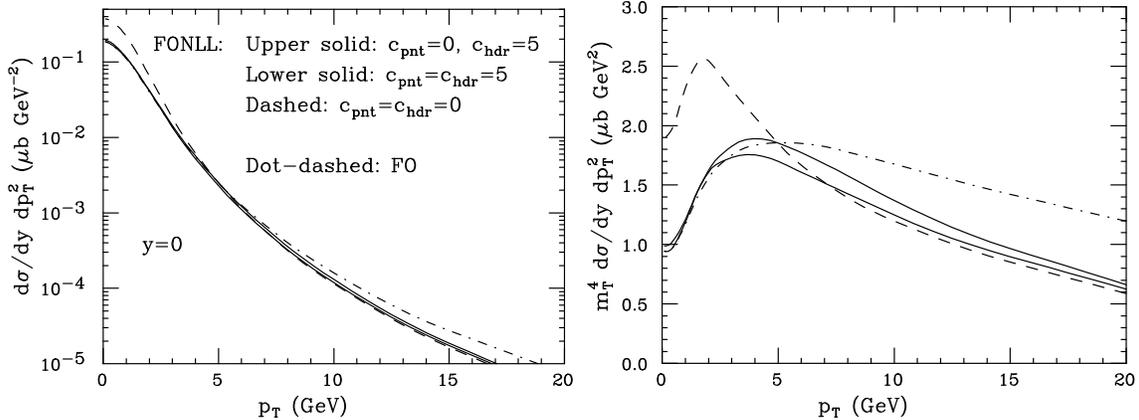

  \begin{center}
    \epsfig{figure=pt-rap0-nosuda-cdcr.eps,height=5.5cm}
    \epsfig{figure=pt-rap0-nosuda-cdcr-mt.eps,height=5.6cm}
\parbox{\capwidth}{
\caption{\label{fig:cdcr} \protect\small
  Effect of the smearing functions in the FONLL result (eq.~(\ref{FONLLOcdcr}))
  for the $\pt$ distribution of charm quarks at $y=0$.
  In the right figure, the same results are multiplied by $\mt^4$.
}}
  \end{center}
\end{figure}

\begin{figure}[tb!]
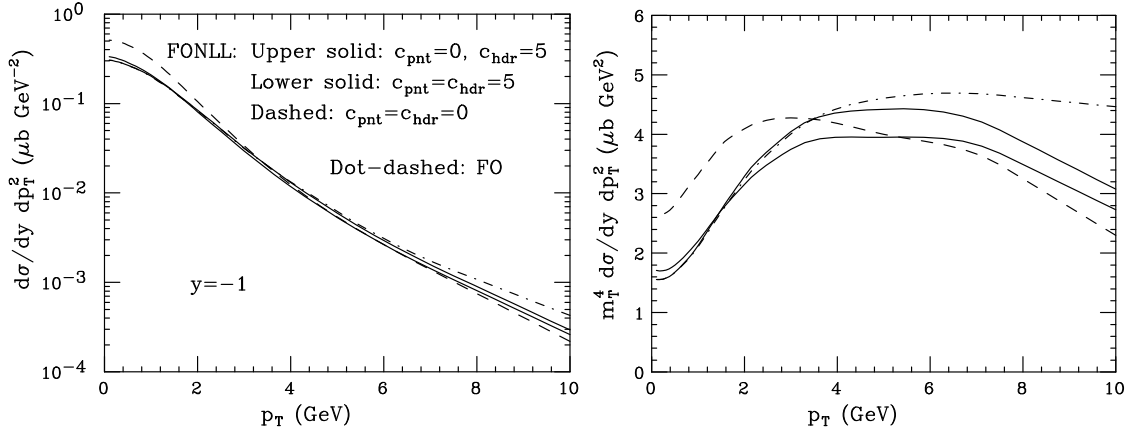

  \begin{center}
    \epsfig{figure=pt-rap-1-nosuda-cdcr.eps,height=5.7cm}
    \epsfig{figure=pt-rap-1-nosuda-cdcr-mt.eps,height=5.6cm}
\parbox{\capwidth}{
\caption{\label{fig:cdcrym1} \protect\small
  As in fig.~\ref{fig:cdcr}, for $y=-1$.
}}
  \end{center}
\end{figure}

At this point, one is tempted to apply the suppression factor $G(m,\pt)$
only to the difference
${\rm RS_{hdr}}-{\rm FOM0_{hdr}}$.
However, this would be incorrect. As we have already discussed in section
\ref{sec:matching}, the terms of order $\aem\as$ and $\aem\as^2$ 
in ${\rm RS_{hdr}}$ do not match ${\rm FOM0_{hdr}}$. The difference
${\rm RS_{hdr}}-{\rm FOM0_{hdr}}$ is of order $\aem\as^2$. Thus, using a
different $c$ value for the pointlike and hadronic component would
lead to the introduction of mass suppressed terms of order
$\aem\as^2$, that would spoil the accuracy of our calculation.
It is instead sensible to use different $c$ values in the following
expression
\begin{eqnarray}
\mbox{FONLL}
&=&{\rm  FO_{pnt} + FO_{hdr} +
 (RS_{pnt}-FOM0_{pnt}+PQ)}\frac{\pt^2}{\pt^2+c_{\rm pnt}^2 m^2} + \nonumber \\
&&{\rm \phantom{\,FO_{pnt} + FO_{hdr} +}
   (RS_{hdr}-FOM0_{hdr}-PQ)}\frac{\pt^2}{\pt^2+c_{\rm hdr}^2 m^2}, 
\label{FONLLOcdcr}
\end{eqnarray}
since now the expressions in parentheses are indeed of the appropriate
order. From formula (\ref{FONLLOcdcr}) with $c_{\rm pnt}=c_{\rm hdr}=c$
we recover the
standard expression eq.~(\ref{eq:merge2}).
In fig.~\ref{fig:cdcr}  we show the effect of varying independently $c_{\rm pnt}$
and $c_{\rm hdr}$ in the FONLL result of eq.~(\ref{FONLLOcdcr}).
It is quite clear that the inclusion of a smearing function for
the pointlike component has little effect (the difference between the two 
solid lines). The smearing function for the hadronic component has 
instead a very large effect,
suppressing the cross section at small $\pt$. In its absence (dashed curve),
one gets a cross section that is roughly twice as large as the fixed order one.
At large $\pt$ the sensitivity to a smearing function decreases rapidly.
The FONLL approach gives smaller cross sections than the fixed order approach in
this region. This is what one expects from resummation effects, since
the emission of collinear gluons, in general, has the effect of softening the
$\pt$ spectrum.
Similar plots are also shown in fig.~\ref{fig:cdcrym1} for $y=-1$.
Also here we see that the effect of the smearing function on the
pointlike component of the cross section is small, and we see a
relatively large effect of the smearing in the hadronic
component. This effect is not quite as large as in the case of $y=0$,
because at negative rapidity the hadronic component is less important.

\begin{figure}[tb!]
  \begin{center}
    \epsfig{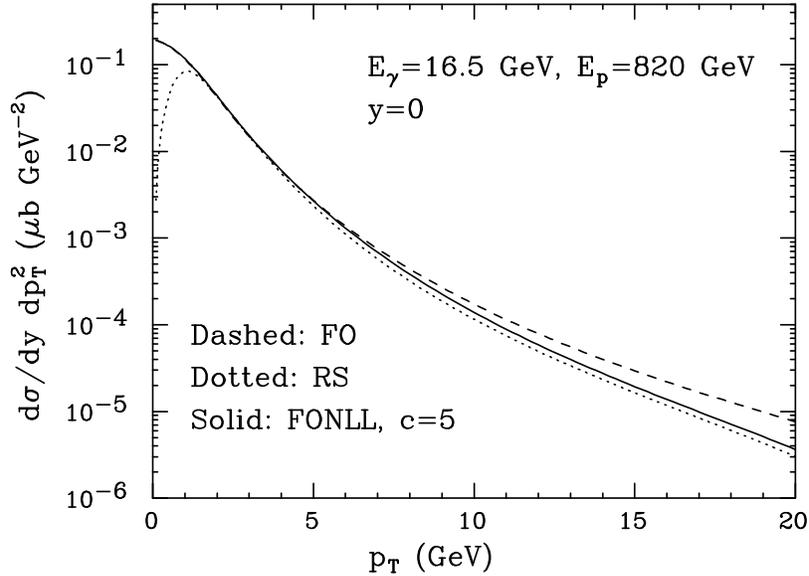}
\parbox{\capwidth}{
\caption{\label{fig:finalp6y0} \protect\small
  Comparison between the FONLL, FO and RS results.
}}
  \end{center}
\end{figure}

\begin{figure}[tb!]
  \begin{center}
    \epsfig{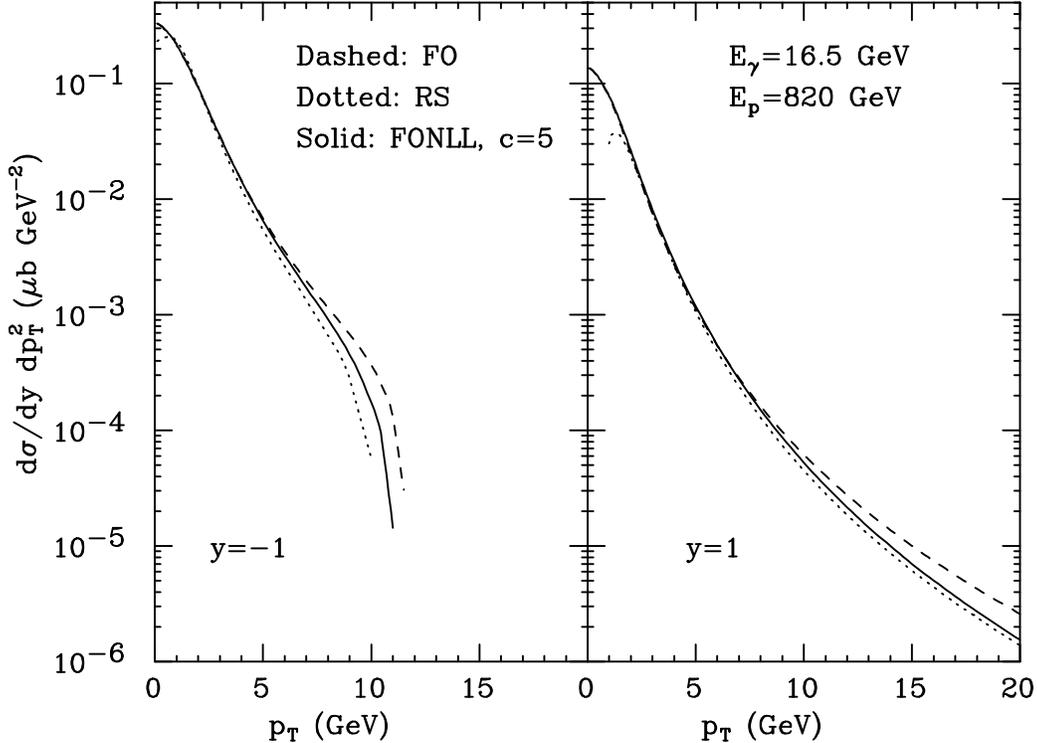}
\parbox{\capwidth}{
\caption{\label{fig:finalp6ypm1} \protect\small
  As in fig.~\ref{fig:finalp6y0}, for $y=-1$ and $y=1$.
}}
  \end{center}
\end{figure}

\begin{figure}[tb!]
  \begin{center}
    \epsfig{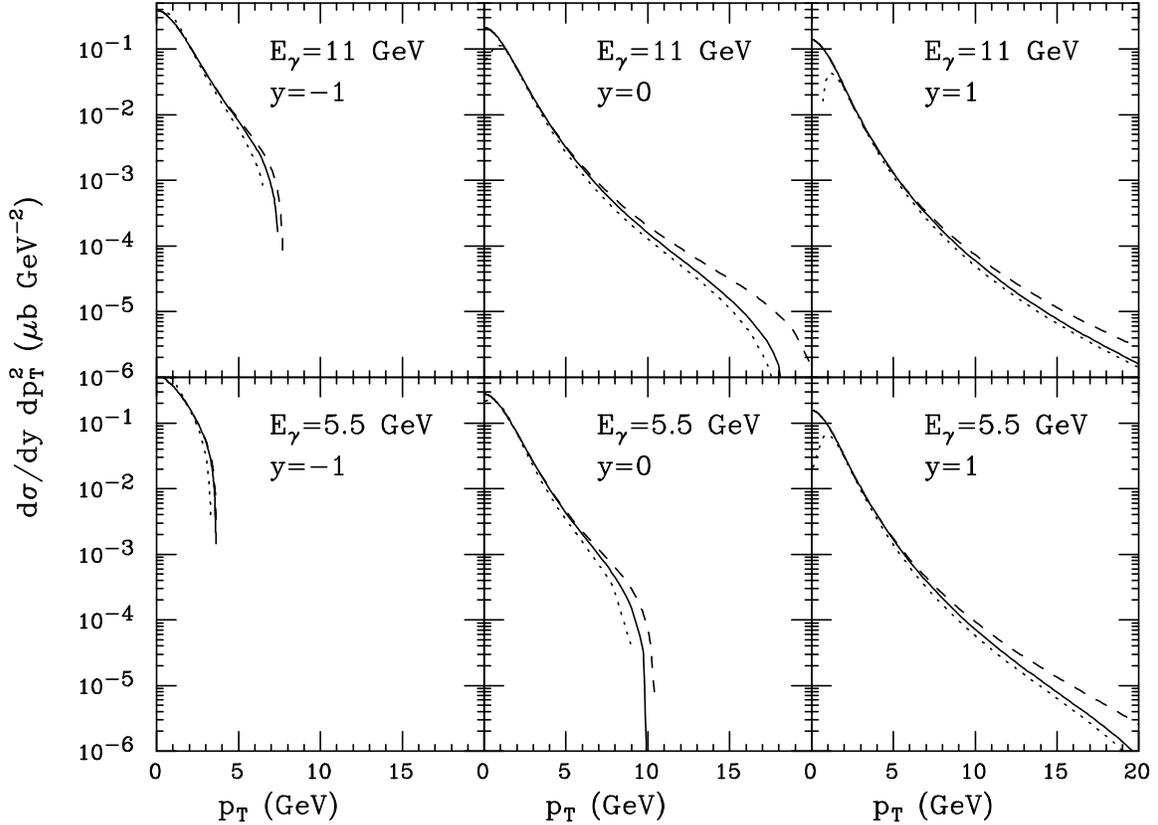}
\parbox{\capwidth}{
\caption{\label{fig:finalp2p4} \protect\small
  As in figs.~\ref{fig:finalp6y0} and \ref{fig:finalp6ypm1}, for
  $E_\gamma=11$ and $5.5$ GeV.
}}
\end{center}
\end{figure}

We shall now show a few benchmark results for photoproduction of heavy flavour
at HERA. For simplicity, we shall always keep
$c\equiv c_{\rm pnt}=c_{\rm hdr}=5$. The aim of these results is to assess the
difference between the fixed order results, the resummed results, and our
matched formalism. We assume an incident photon energy of 16.5~GeV
(corresponding to an electron energy of 27.5 GeV and to a photon energy
fraction of 0.6), and a proton energy of 820 GeV. In fig.~\ref{fig:finalp6y0}
we show a comparison between the FONLL, FO and RS results. We see that
the RS and FO results are remarkably close for small $\pt$, a fact
that we have understood to be true for the pointlike component,
and purely accidental for the hadronic component. At moderate $\pt$ also
the FONLL result is very close to the FO result. While in the case
of the pointlike component this is true in all cases, for the hadronic
component this only happens if one chooses a relatively large value for $c$.
At very small $\pt$ the resummed calculation becomes totally unreliable.
At large $\pt$, the FONLL and RS results become quite similar (since, by
construction, the $G(m,\pt)$ suppression factor tends to one, and FO tends to
FOM0, canceling it - see eqs.~(\ref{eq:merge}) and (\ref{eq:merge2})), 
and remain
smaller than the FO result. This is a general consequence of the effect
of multi-gluon emission resummed in the RS and FONLL approaches.
The pattern displayed in fig.~\ref{fig:finalp6y0} seems to be quite universal.
Similar plots for $y=1$ and $y=-1$ are shown in
fig.~\ref{fig:finalp6ypm1}.
Furthermore in fig.~\ref{fig:finalp2p4} we also show the case of photon energy
fractions of 0.4 and 0.2 (that is $E_\gamma=11$ and $5.5$ GeV respectively).

\section{Conclusions} \label{sec:conclusions}
In this work we have implemented a technique to compute the transverse
momentum spectrum in heavy flavour photoproduction, which is accurate
to the full NLO level at moderate $\pt$ values, and to the NLL level
at large $\pt$. 

This is achieved by properly merging a ``massless'' resummed approach, valid in
the large-$\pt$ region, with a full massive fixed order calculation, reliable in
the small-$\pt$ one.

We observe that, for the pointlike component of the cross section (in the
sense discussed in section~\ref{sec:matching}), the massless limit is a good
approximation to the full cross section, provided one uses $\mt$ rather than
$\pt$ in the former. On the contrary, this is not the case for the hadronic 
component of the cross section, as already observed in ref.~\cite{cgn}
in the hadroproduction case: power suppressed mass terms happen to be important
in the moderate $\pt$ region. This suggests that a proper merging like the one
studied in this paper is necessary and superior to just employing a massless 
approach.

The results obtained with our procedure are in good
agreement with the fixed-order calculations at moderate transverse momenta,
and with the so-called massless resummation (or fragmentation function) 
approach at very large momenta, effectively interpolating, in a theoretically
sound manner, between the two approaches.

In general, we find that inclusion of resummation effects brings about
a softening of the $\pt$ spectrum at large transverse momenta.

\vspace{.7cm}
\noindent
{\bf Acknowledgments.} The authors thank the CERN Theory Division, where much
of this work was performed. The work of M.C. was supported in part by the
National Science Foundation Contract PHY-9722101.

\end{document}